\documentclass{article}

\usepackage{arxiv}

\usepackage[utf8]{inputenc} 
\usepackage[T1]{fontenc}    
\usepackage{hyperref}       
\usepackage{url}            
\usepackage{booktabs}       
\usepackage{amsfonts}       
\usepackage{nicefrac}       
\usepackage{microtype}      
\usepackage{times}
\usepackage{epsfig}
\usepackage{graphicx}
\usepackage{amsmath}
\usepackage{amssymb}
\DeclareRobustCommand\onedot{\futurelet\@let@token\@onedot}
\def\@onedot{\ifx\@let@token.\else.\null\fi\xspace}

\def\etc{\emph{etc}\onedot} 
 
\def\etal{\emph{et al}\onedot}

\title{Light Field Compression by Residual CNN Assisted JPEG}

\author{Eisa Hedayati\\
Michigan Technological University\\
{\tt\small hedayati@mtu.edu}
\And
Timothy C. Havens\\
Michigan Technological University\\
{\tt\small thavens@mtu.edu}
\And
Jeremy P. Bos\\
Michigan Technological University\\
{\tt\small jpbos@mtu.edu}
}

\begin{document}

\maketitle

\begin{abstract}
   Light field (LF) imaging has gained significant attention due to its recent success in 3-dimensional (3D) displaying and rendering as well as augmented and virtual reality usage. Because of the two extra dimensions, LFs are much larger than conventional images. We develop a JPEG-assisted learning-based technique to reconstruct an LF from a JPEG bitstream with a bit per pixel ratio of 0.0047 on average. For compression, we keep the LF's center view and use JPEG compression with 50\% quality. Our reconstruction pipeline consists of a small JPEG enhancement network (JPEG-Hance), a depth estimation network (Depth-Net), followed by view synthesizing by warping the enhanced center view. Our pipeline is significantly faster than using video compression on pseudo-sequences extracted from an LF, both in compression and decompression, while maintaining effective performance. We show that with a 1\% compression time cost and 18x speedup for decompression, our methods reconstructed LFs have better structural similarity index metric (SSIM) and comparable peak signal-to-noise ratio (PSNR) compared to the state-of-the-art video compression techniques used to compress LFs.
\end{abstract}

\section{Introduction}

\emph{Light fields} (LF), as compared to conventional images, have two extra dimensions which represent angular information of the scene \cite{adelson1991plenoptic,Levoy1996,ng2005}. Hence, LFs contain a relatively large volume of data that makes storing and portability time consuming and costly. Also, decompression of LF video with a high angular resolution at acceptable \emph{frames per second} (fps) for streaming is challenging. We aim to address these challenges by predicting the entire LF from its JPEG compressed center view. 

Direct application of standard image compression techniques, such as JPEG, PNG, \etc, on an LF does not take advantage of existing redundancies between LF views. Video compression techniques, however, achieved better success in compressing LFs. To use video compression algorithms on LFs, a sequence of images is built from LF views, which is called pseudo-sequence\cite{Liu2016}. A combination of \emph{machine learning} (ML) methods, capable of predicting LF views, and video compression techniques was explored in \cite{Z.Zhao2018}. In this work, we present a combination of JPEG compression with ML view predictions. LF synthesis techniques have shown the possibility of estimating the entire LF from a single view or a set of sparse views. Here, we show that there is enough information in the JPEG compressed center view---as well as a group of \emph{sub-aperture images} (SAIs)---to predict the entire LF with a quality comparable to the use of the state of the art video compression techniques on the LF. We test the success of our method by comparing against state-of-the-art methods in LF compression that use the existing HEVC compression.

Our method is faster in compression and decompression by 100x and 10x, respectively, compared to the direct use of HEVC. This speed up means a set of 30 LFs with a spatial resolution of $(375,540)$ and angular resolution of $(7,7)$ can be decompressed on a typical gaming GPU in less than 0.02 seconds, while HEVC-based methods require more than 0.39 seconds. With increases in spatial or angular dimensions, HEVC-based methods reconstruction soon takes more than one second. Thus, streaming will not be possible without pre-decompression. Furthermore, while speeding up the process, we have maintained and, in most cases, improved the quality of reconstruction at the same \emph{bit per pixel} (bpp) ratio.  We have used the \emph{mean of peak signal-to-noise} (MPSNR) ratio over all of the views  and \emph{mean structural similarity index metric} (MSSIM) to compare the reconstructed LFs of our model with those that use HEVC. We show that while the MPSNR of our method is comparable to the direct employment of HEVC, our model can achieve higher MSSIM. This results in fewer artifacts and better quality in the extracted synthetic aperture \emph{depth of field} (DoF) images. Note that we built our model to be fully convolutional, thus, it can be used on LFs with any spatial resolution. Also, our model works in the RGB channel. This is an advantage compared to other techniques, which are using YUV channel, because most of the available LF datasets are in RGB and conversion between RGB and YUV is not lossless.

The contributions of this paper are as follows. We achieved compression speed-up of more than 100x and decompression speed-up of more than 10x compared to the use of HEVC on pseudo-sequences of LF views. At an average bpp of $.0047$, the LF's DoF reconstructed with our method improved the SSIM by $0.31\%$ on average over the test dataset compared to the direct use of HEVC. Finally, we introduce a small, fast, and efficient \emph{convolutional neural network} (CNN) for enhancing JPEG images for use in LFs. This network also boosts the SSIM of the final decompressed LF.

\section{Related Work}
\subsection{Light Field View Synthesis}
Linear view synthesis by Levin and Durand \cite{Levin2010} and depth of field extension and super-resolution by Bishop and Favaro \cite{Bishop2012} are among the earliest works on LF view synthesis and reconstruction. Flynn \etal \cite{Flynn_2016_CVPR} proposed a deep learning method to predict novel views from a sequence of images with wide baselines. LF view synthesis became more popular after Kalantari \etal \cite{Kalantari2016} showed in their work that an LF can be synthesized from its corner SAIs with high quality. Building on the work of Kalantari \etal, Yeung \etal used different sets of views to reconstruct dense LFs \cite{Yeung_2018_ECCV}. Srinivasan \etal demonstrate the possibility of estimating the entire LF from its center view by extrapolating using machine learning methods \cite{Srinivasan_2017_ICCV}. Choi \etal extended the extrapolation to an LF taken with arrays of cameras \cite{choi2019}. LF fusion \cite{mildenhall2019} and depth-guided techniques \cite{WenhuiZhou2020} have been popular in reconstructing an LF from a single or a sparse set of SAIs.
Hu \etal \cite{ZexiHu2020} aimed for a faster LF reconstruction method by using hierarchical features fusion.

The backbone of nearly every view synthesis method enumerated here is the depth-map estimation.
The current work is categorized as single view LF reconstruction. Our method is unique because we use a lossy compressed JPEG view from which to estimate the entire LF. We use residual learning methods to guess the possible artifacts from the JPEG compressed version of the center view to assist the main network for accurately estimating the depth map. 
\subsection{Light Field Compression}
Lossless and lossy compression methods have been investigated extensively in the literature. For the lossless model, Perra \cite{perra2015lossless} proposed an adaptive block differential prediction method and Helin \etal \cite{helin2016sparse} described a sparse modeling with a predictive coding for SAIs of the LF.

The lossy models can be classified in to sub-categories of: i) standardized image/video compression techniques and ii) machine learning assisted compression techniques.
\subsubsection{LF compression by standardized image/video compression methods}

Standardized image and video compression techniques (especially HEVC) have been directly used to address the problem of the bulkiness of the LF, see e.g., \cite{Liu2016,Conti2016,Li2016}. Other methods, such as homography-based low-rank models \cite{XiaoranJiang2017} and Fourier disparity layers \cite{Dib2019}, have been used to reduce the angular dimension of the LF. In another work, the LF depth was segmented into 4D spatial-angular blocks, which were used for prediction, followed by encoding the residue using JPEG-2000 \cite{Tabus2017}.

\subsubsection{Machine learning assisted compression techniques}
Followed by the breakthrough in synthesizing LF views from its four corners using CNN learning techniques introduced by \cite{Kalantari2016}, another work introduced a compression technique by using the same method and compressing the four corner views by HEVC \cite{Jiang2017}. In another work, the authors proposed to keep half of the views and encode them by HEVC and synthesize the other half by a CNN \cite{Z.Zhao2018}. A CNN based epipolar plane image super-resolution algorithm was used in cooperation with HEVC to compress LF as well \cite{JZhao2018}.
Wang \etal proposed a new LF video compression technique by deploying view synthesis methods from multiple inputs while encoding the input views by a proposed region-of-interest scheme \cite{B.Wang2019}. Generative adversarial network based methods have been used in cooperation with video codec techniques to compress LF in \cite{Gan2019,Gan2021}. There has been multiple works on LF compression with use of standard video compression codecs in combination with learning based view synthesis \cite{bilevel2019,jWang2020}. Unfortunately, we could not find any public version of these codes, or trained networks for the purpose of comparing our results with them. Hence, we have chosen the pseudo-sequence HEVC compression method for comparison because of its easy implementation and availability of the HEVC codec.

To the best of our knowledge, because the view extrapolation is ill posed, LF reconstruction from a lossy compressed single input (specifically, JPEG) has not been explored before our work.

\subsection{JPEG Compression Artifact Reduction}
For several decades, different researchers addressed the JPEG compression artifact reduction generally in three main groups: prior knowledge-based, filter-based, and learning-based approaches. Here though, we are interested in learning-based approaches. The basic intention of learning-based methods is to find a non-linear mapping between the JPEG compressed image---compressed at different compression ratios---to the ground truth uncompressed image. To the best of our knowledge, the first deep learning model to address this problem was created by Dong \etal \cite{Dong_2015_ICCV}, where they showed the possibility of enhancing the reconstructed JPEG image by a relatively shallow CNN. Since then, multiple researchers have gradually improved the performance of learning-based methods by introducing new networks such as: dual-domain representations \cite{JGuo2016}, deep dual-domain based fast restoration  \cite{Wang_2016_CVPR}, encoder-decoder networks with symmetric skip connection \cite{MaoNIPS2016}, CAS-CNN \cite{Cavigelli2017}, one-to-one networks \cite{BaochangZhang2018}, DMCNN \cite{XiaoshuaiZhang2018}, and dual-stream multi-path recursive residual network \cite{ZhiJin2020}. While deeper networks and state of the art architectures have improved the task of JPEG artifact reduction, we are not focused solely on this task here. The ultimate goal of our JPEG-Hance network is to improve the estimated depth-map from the JPEG compressed center image of an LF. JPEG artifact reduction is the natural first step for extracting better depth-maps. 

\section{The Proposed Method}
Here we describe our compression and decompression pipeline. The compression pipeline is simply extraction of the center view of the LF, compression by JPEG at 50\% quality, followed by discarding of all other views. The decompression pipeline has the following steps:
\begin{enumerate}
	\item JPEG decompression of the center view $c_J$
	\item Enhancing $c_J$ by JPEG-Hance to $c_E$,
	\begin{equation}\label{eq:J-H}
		c_E = J(c_J).
	\end{equation}
	\item Estimating depth map $d(x,u)$ of every view $u$ from $c_E$,
	\begin{equation}\label{eq:D-M}
		d = D(c_E).
	\end{equation}
	\item Reconstructing LF by
	\begin{equation}\label{eq:lamWarp}
		\mathit{L(x,u)}_{u_0\rightarrow u} = L(x+(u-u_0)d(x,u),u_0),  
	\end{equation}
	where $L$ is the approximated LF and $u_0$ is the middle view index. Variables $x$ and $u$ are spatial and angular indices.
\end{enumerate}
\subsection{Networks architectures}
\subsubsection{JPEG-Hance}
\begin{figure}
	\centering
	\includegraphics[width=0.45\textwidth]{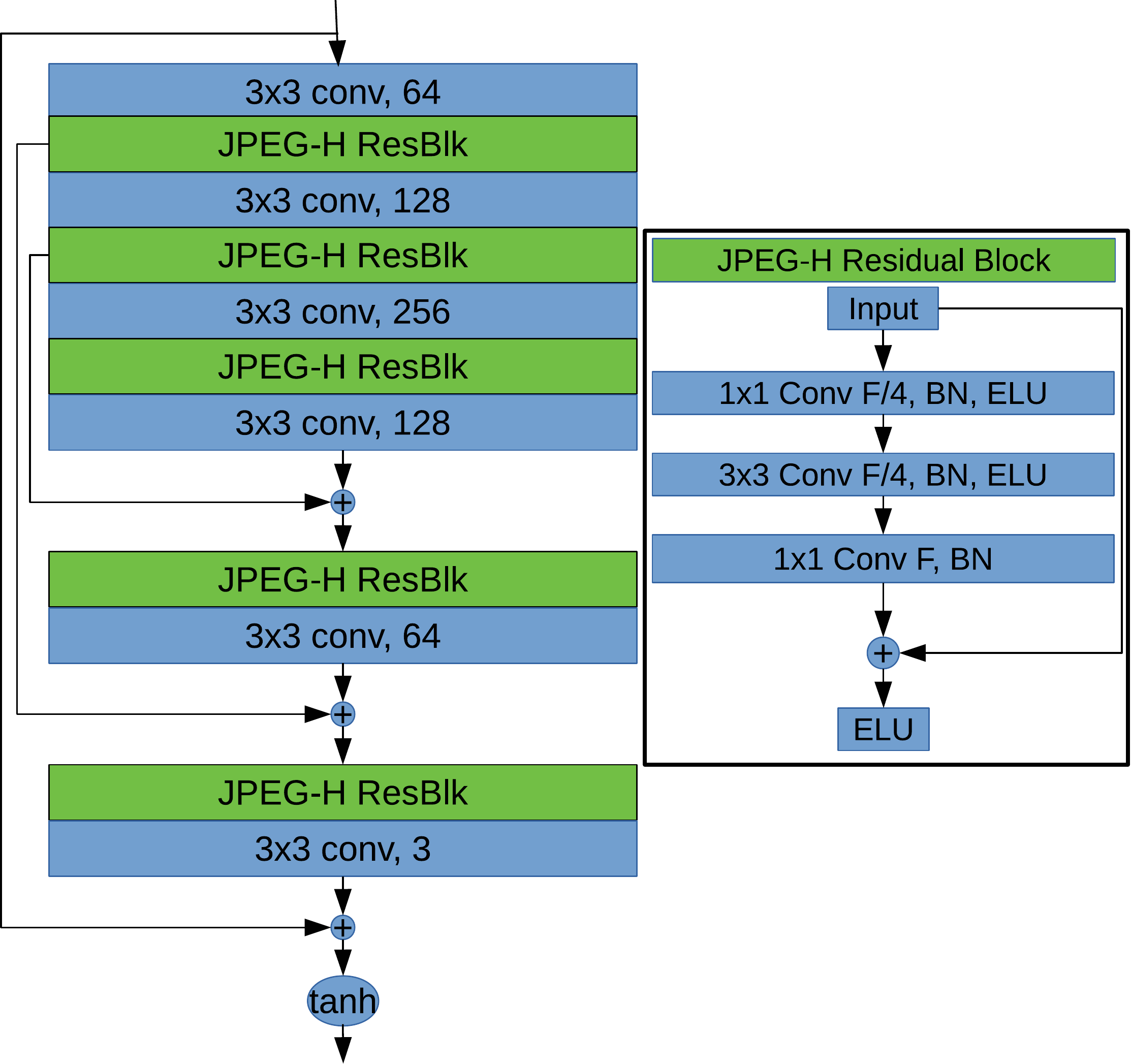}
	\caption{\label{fig:J-H}JPEG-Hance detailed structure}
\end{figure}

The main goal of our JPEG-Hance network is to assist the Depth-Net in providing better depth map estimation. In doing so, it is certainly beneficial to improve the overall quality of the JPEG decompressed image by reducing the error between uncompressed ground truth images and the lossy compressed ones. However, the goal of our network is not general JPEG artifact reduction; instead, JPEG-Hance should learn to enhance the parts of the image which have the most effect on improving depth information extraction. To achieve this task, JPEG-Hance also needs to find correspondence information from the extracted depth maps. Therefore, it is trained in two phases: first, it learns to enhance any typical JPEG decompressed image, then again as part of the whole depth estimation pipeline. 
The architecture of JPEG-Hance is shown in Fig.~\ref{fig:J-H}. Inspired by ResNet50's \emph{bottleneck} building blocks structure \cite{Resnet}, we have designed our JPEG-Hance as residual blocks. We added a \emph{batch normalization} (BN) layer after each convolution followed by an \emph{exponential linear unit} (ELU). The \emph{ELU} followed by a last layer \emph{tanh} seems to be the most promising activation pair of functions when dealing with regression of image data scaled to the interval $[-1,1]$.

JPEG-Hance is pre-trained by minimizing the mean squared error of each pixel value in the RGB channels. Then it is added to the training pipeline for full reconstruction of LFs.
\subsubsection{Depth-Net}
Multiple images provide geometry information which can be used for LF reconstruction. A single image does not provide such information. Therefore, such information needs to be extracted by other methods. Machine learning techniques, particularly CNNs, showed a promising potential for estimating geometry from a single image \cite{Srinivasan_2017_ICCV,Kalantari2016,Yeung_2018_ECCV}. Thus, for the problem of depth estimation from our enhanced center image, we use a residual CNN.

\begin{figure}
	\centering
	\includegraphics[width=0.45\textwidth]{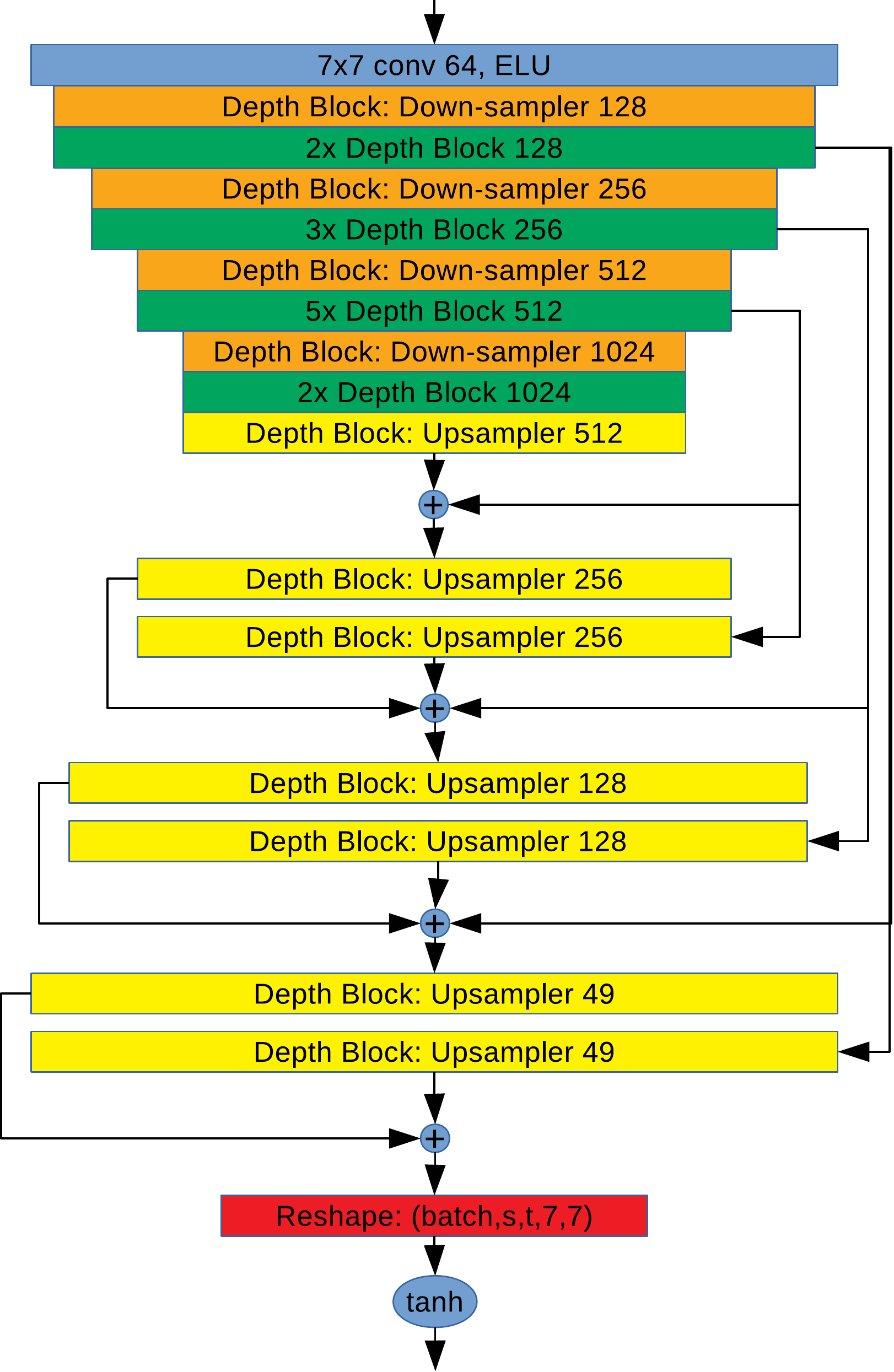}
	\caption{\label{fig:D-N}Depth-Net detailed structure}
\end{figure}

\begin{figure*}
	\centering
	\includegraphics[width=.85\textwidth]{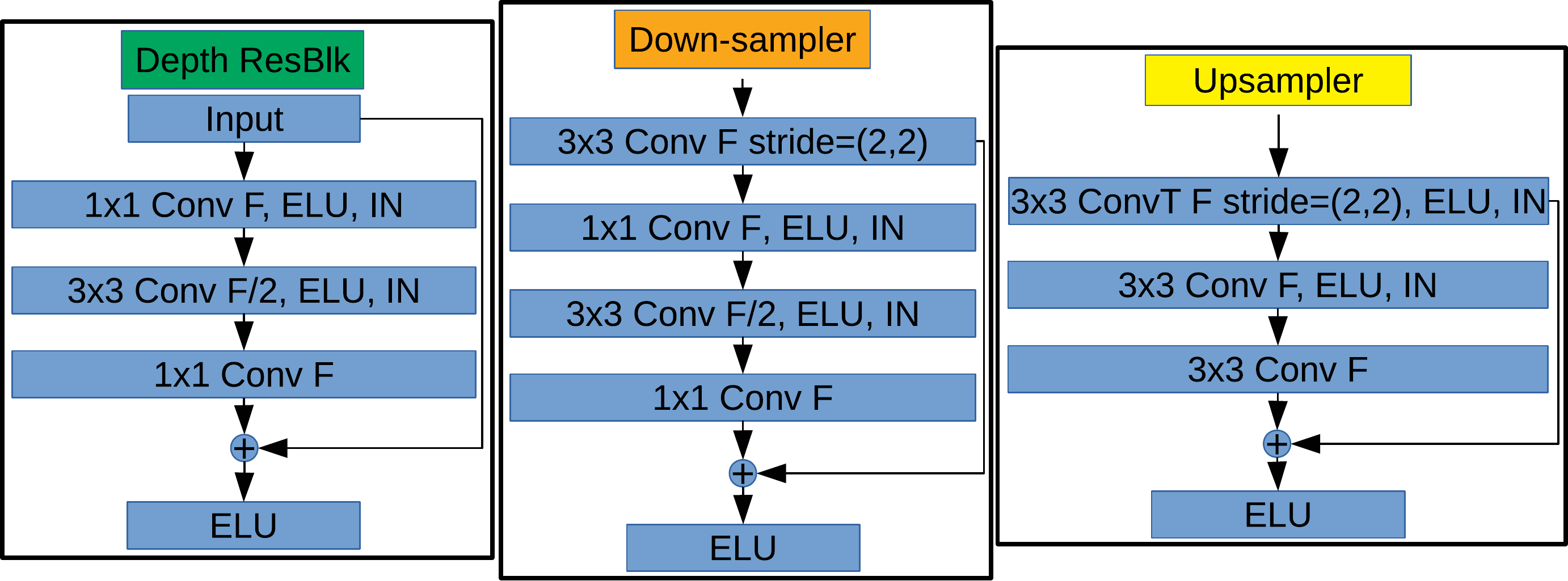}
	\caption{\label{fig:D-N-B}Depth-Net residual blocks}
\end{figure*}

Our Depth-Net, depicted in Fig.~\ref{fig:D-N}, is responsible for estimating the depth map (disparity map) for all 49 views from the middle JPEG compressed view. Depth-net has three variants of residual blocks. The first variant is a down-sampler which uses a 2D convolution with strides of $(2\times 2)$, halving the spatial dimension of the input image. This block is used just before the first Depth Residual Block; each time the feature size is increased. The second type of block, the Depth Residual block, is the main residual block. This block is used the most and extracts most of the features. The structure of the Depth Residual block mimics the bottleneck structure of ResNet50 with added instance normalization after each of the first two convolution layers. Last, the Upsampler block is constructed to have a 2D deconvolution (transposed convolution) layer and two 2D convolution layers with kernel size of $(3\times3)$. The deconvolution layer's stride is set to $(2\times2)$. These blocks are shown in Fig.~\ref{fig:D-N-B}.

Because we are training the Depth-Net on the actual LF data and not the ground truth depth maps, our loss functions have to be designed to train the network in an unsupervised manner. Our Depth-net is predicting the LF's depth while we do not have the ground truth depth to supervise the training. Thus, we define the Depth-Net pre-training loss function to be a weighted sum of four sub-functions: i) photometric loss $\mathit{L_p}$, ii) defocus loss $\mathit{L_r}$, iii) depth-consistency loss $\mathit{L_c}$, and iv) DoF loss $\mathit{L_d}$. The total loss is simply
\begin{equation}\label{eq:depthLoss}
	\mathit{L}_{depth} = \alpha \mathit{L}_p + \alpha_1 \mathit{L}_r + \alpha_2 \mathit{L}_c + \alpha_3 \mathit{L}_d,
\end{equation}
where $\alpha,\alpha_1,\alpha_2$ and $\alpha_3$ are chosen to be $2$, $100$, $0.02$, and $10$ in our conducted experiments, which were empirically found to work well overall.

The image quality comparison sub-function $\psi$ \cite{WenhuiZhou2020} is constructed by combining mean absolute difference of pixels and image structural dissimilarity (DSSIM) that is derived from the \emph{structural similarity index} metric (SSIM) \cite{SSIM}:
\begin{equation}\label{eq:psi}
	\psi(I_1,I_2) = \beta\frac{1-SSIM(I_1,I_2)}{2} + (1-\beta)\left\|I_1-I_2\right\|_1,
\end{equation}
where $I_1,I_2$ are two images that are being compared and $\beta\in [0,1]$, which we empirically found that $0.15$ yields the best training results. Using the sub-function $\psi$, photometric loss is defined as \cite{WenhuiZhou2020}
\begin{align} 
	\nonumber    L_p = & \sum_u
	\big[\psi(L(x,u)_{u_0\rightarrow u},L(x,u)) +\\
	\label{eq:photometric}    & \psi(L(x,u)_{u\rightarrow u_0},L(x,u_0))\big]. 
\end{align}

Because we are training an unsupervised Depth-Net, the more prior knowledge we can give the network, the better will be the training quality. Zhou \etal \cite{WenhuiZhou2020} introduced defocus cue loss
\begin{equation} \label{eq:defocus}
	L_d=\psi\big(L(x,u_0),\frac{1}{N}\sum_u L(x,u)_{u\rightarrow u_0}\big).
\end{equation}
Also depth consistency (left-right or forward-backward) has been shown in the literature \cite{Godard_2017_CVPR,Wang_2018_CVPR,Yin_2018_CVPR} to be a promising regularizer for LF view synthesis purpose, where
\begin{subequations}\label{eq:consistency}
	\begin{align} 
		d_{u_0\rightarrow u}(x) & = d_{u_0}\big(x,(u-u_0)d(x,u)\big), \\
		L_c & = \sum_u ||d_u(x) - d_{u_0\rightarrow u}(x)||_1.
\end{align}\end{subequations}

Finally we have included \emph{depth of field} (DoF) loss to further assist the network in learning depth information, where
\begin{subequations}\label{eq:dofLoss}
	\begin{align} 
		DoF & = \frac{1}{u}\sum_u L(x,u),\\
		L_{dof} & = \psi\big(DoF,DoF_{u_0\rightarrow u}\big).
\end{align}\end{subequations}
\section{Experiments}
In this section, we describe our method's implementation details. Then, we use public data sets \cite{Srinivasan_2017_ICCV,Kalantari2016} to evaluate our method and investigate the impact of different parts of our network on the performance of our model.

\subsection{Data sets}
We have conducted our experiments over the two public data sets: \emph{Flowers} \cite{Srinivasan_2017_ICCV} and \emph{30 Scenes} \cite{Kalantari2016}. Both of these data sets are captured by a Lytro Illum camera. The angular resolution of these data sets is $14\times14$ views and the spatial resolution is variable between $375\times540$ and $376\times541$ pixels. The LF from these data sets are cropped to the size of $7\times7\times375\times540$ to have a consistent size and vignetting.

\subsection{Implementation details}
Our pipeline was trained in multiple steps. We have implemented our model with Tensorflow 2.2 in python 3.7 on a workstation with an Intel Xeon W-2223 3.60 GHz, 64GB DDR4 memory, and NVIDIA Quadro RTX 5000.

\subsubsection{JPEG-Hance}
Our JPEG-Hance was pre-trained on the \emph{30 Scenes} training data set, which contains 100 scenes. The center views of these 100 scenes were extracted and used for training. In the training phase, the spatial dimensions of the JPEG-Hance were set to $128\times128$. First a training pool of images with dimensions of $128\times128$ was created by cropping the center views of the 100 scenes at 8 pixels steps. Therefore, the training pool had 150,000 different crops which we found sufficient for the JPEG-Hance network to be trained without over-fitting or under-fitting. The learning rate was set to 0.0004. The JPEG-Hance has a relatively small network: only 202,435 trainable parameters. The pre-training phase took about 90 minutes to converge.

\subsubsection{Depth-Net}
Our Depth-Net also has a pre-training step. The Depth-Net was pre-trained on the \emph{Flowers} data set, which has 3,343 flowers. During the pre-training phase, the JPEG-Hance was used for enhancing center images while only Depth-Net parameters were trained. The input pipeline of the flowers contains random croppings to $128\time128$ and data augmentation with 50\% selection rate for the original data, 15\% chance for random contrast change between $[0.1,0.5]$, 15\% chance that the brightness was changed randomly up to $0.4\times$ original brightness, and the remaining 20\% where the hue was randomly changed by up to $0.4\times$. During the pre-training phase, 16 random crops were extracted for each epoch, and the network was trained for 10 epochs. The learning rate was 0.0004. Depth-Net is the main network responsible for extracting the depth map; thus, it has more trainable parameters: about 38.2 million. The pre-training phase takes about 7 hours to converge.

\begin{figure}
	\centering
	\includegraphics[width=.45\textwidth]{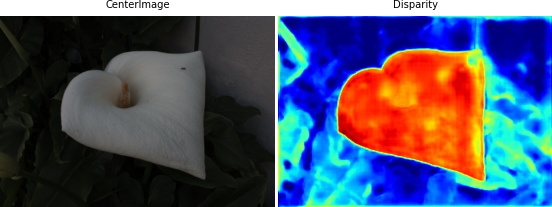}
	\caption{\label{fig:disp} An estimated depth map. We can see that our network estimate is correct for most parts of the image. The map indicates that the flower is the nearest object to the LF camera and the leaves are just behind the flower and the wall is at the background, which is very realistic. }
\end{figure}

A sample estimated depth is depicted in Fig.~\ref{fig:disp}. This illustration demonstrates that the edges are not very sharp. This is because we have used the highly compressed lossy JPEG on center view, which adds blur to the edges, to estimate the disparity map. Thus, the resulting depth map is somewhat blurry.

\subsubsection{Training the entire pipeline}
After pre-training the two networks, we can now train the entire pipeline. We add 100 scenes from the \emph{Flowers} data set pool and use the same input pipeline as the one used for Depth-Net. The entire pipeline was trained for 45 epochs, gradually decreasing the learning rate from 0.0001 to 0.000001 for the last 5 epochs.

The last fine-tuning step includes training the pipeline on the data sets with input spatial dimensions of $375\times540$. Here the augmentation selection is 25\% original, 25\% random contrast, 25\% random brightness, and 25\% random hue. Because of the structure of the Depth-Net network, the input images have to be zero-padded and the resulting LFs should be cropped to the correct size. The input dimension of the Depth-Net is $384\times544$. The fine tuning phase takes 40 epochs for the network to converge with gradually decaying learning rate from 0.00005 to 0.000001. The fine-tuning phase took around 20 hours to converge, while all other pre-training phases took less than 10 hours cumulatively.

\subsection{Performance comparison}
We compared our compression-decompression results with a pseudo sequence method using the HEVC video compression Codec. We chose a raster sequence over spiral because raster had slightly better performance.
The 30 scenes data set was used for comparing our method with HEVC. We use MSSIM and MPSNR metrics as well as SSIM and PSNR of the extracted DoF from LFs to compare the results, where
\begin{align}
	MSSIM & = \frac{1}{M} \sum_u{SSIM(LF,\Tilde{LF})},\\
	MPSNR & = \frac{1}{M} \sum_u{PSNR(LF,\Tilde{LF})}.
\end{align}

\begin{figure}
	\centering
	\includegraphics[width=.45\textwidth]{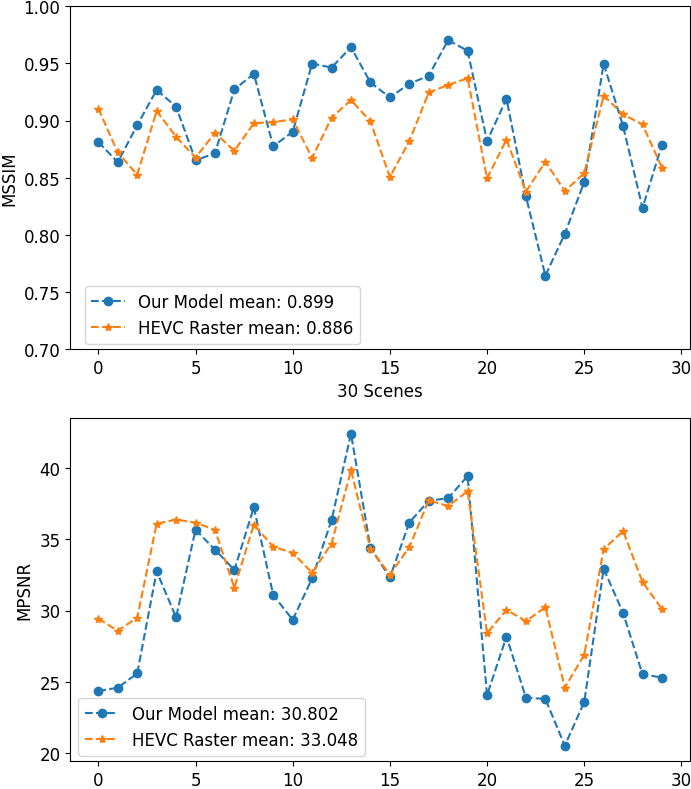}
	\caption{\label{fig:MSSIMPSNR} The top figure is showing the MSSIM for each reconstructed LF by our method and HEVC. The bottom figure is the MPSNR calculated for each reconstructed LF.}
\end{figure}

\begin{figure}
	\centering
	\includegraphics[width=.45\textwidth]{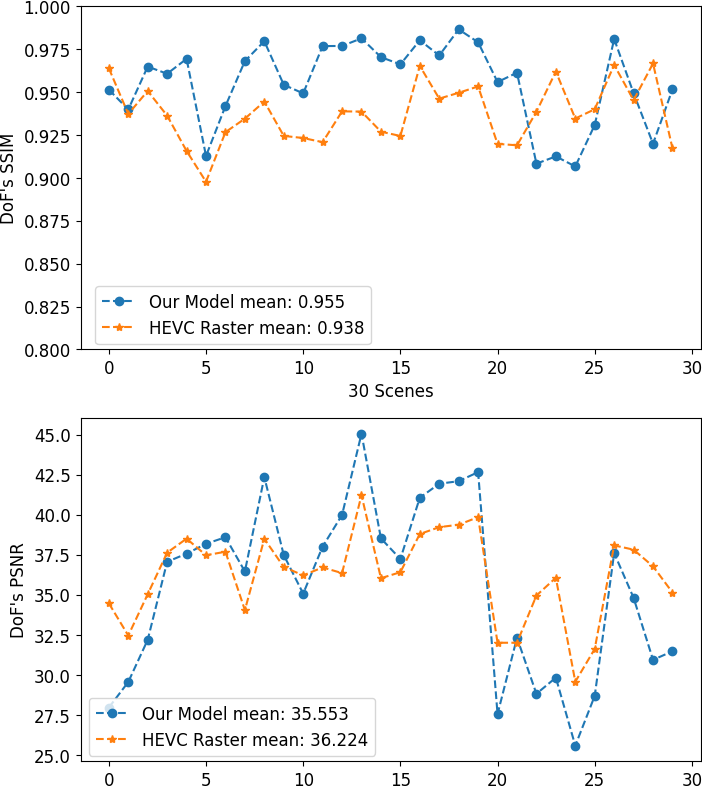}
	\caption{\label{fig:DOFSSIMPSNR} The DoF images extracted from ground truth LFs and the reconstructed LFs are compared using SSIM and PSNR metrics. The top plot is representing SSIM and the bottom one is showing PSNR for each LF in the 30 scenes.}
\end{figure}

To have a fair comparison, we tuned the QP factor of HEVC for each LF to reach approximately similar bpp between HEVC compressed LF and our method's compressed representative. The average bpp for both methods on the 30 scenes data is $0.0047$. Fig.~\ref{fig:MSSIMPSNR} shows that the LFs reconstructed by our method have very similar MPSNR and MSSIM to those decompressed by HEVC. By carefully examining Fig.~\ref{fig:MSSIMPSNR} it is evident that, while our method outperforms HEVC in MSSIM, it is slightly inferior in MPSNR performance.  Fig.~\ref{fig:DOFSSIMPSNR} plots the PSNR and SSIM metrics for extracted DoFs from the reconstructed LFs. Here, our method meaningfully outperforms HEVC in SSIM metrics while further reducing the gap in PSNR. Because of this dual behavior in SSIM and PSNR metrics between our method's results and HEVC's, we have conducted experiment on the refocused images to find out which method is more reliable.

\begin{figure}
	\centering
	\includegraphics[width=.45\textwidth]{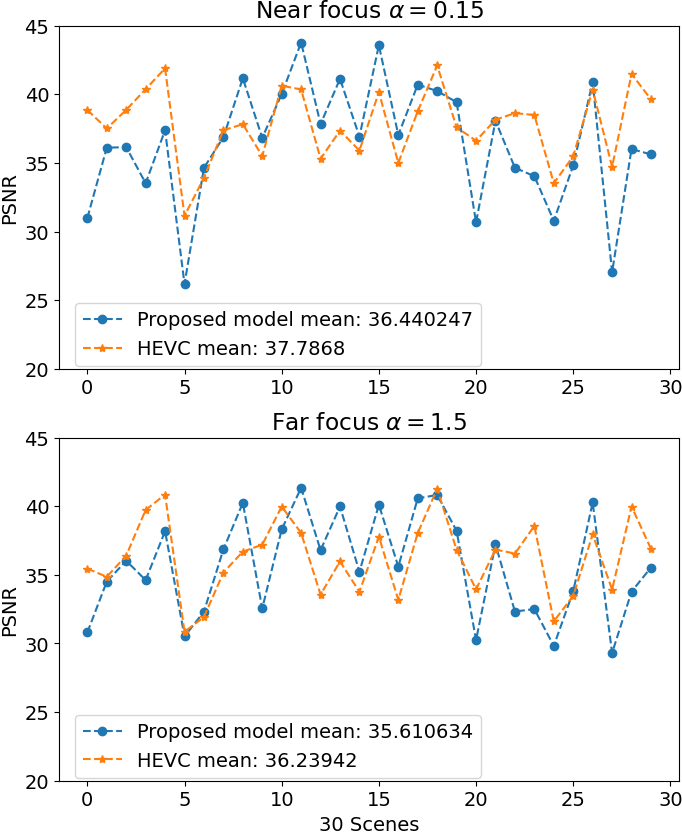}
	\caption{\label{fig:REFPSNR} The reconstructed LFs from our model and HEVC is used to extract refocused images with refocusing parameters of $\alpha=0.15,1.5$, The top plot is showing the PSNR of the near focus images, and the bottom on is the far focus PSNRs.}
\end{figure}

\begin{figure}
	\centering
	\includegraphics[width=.45\textwidth]{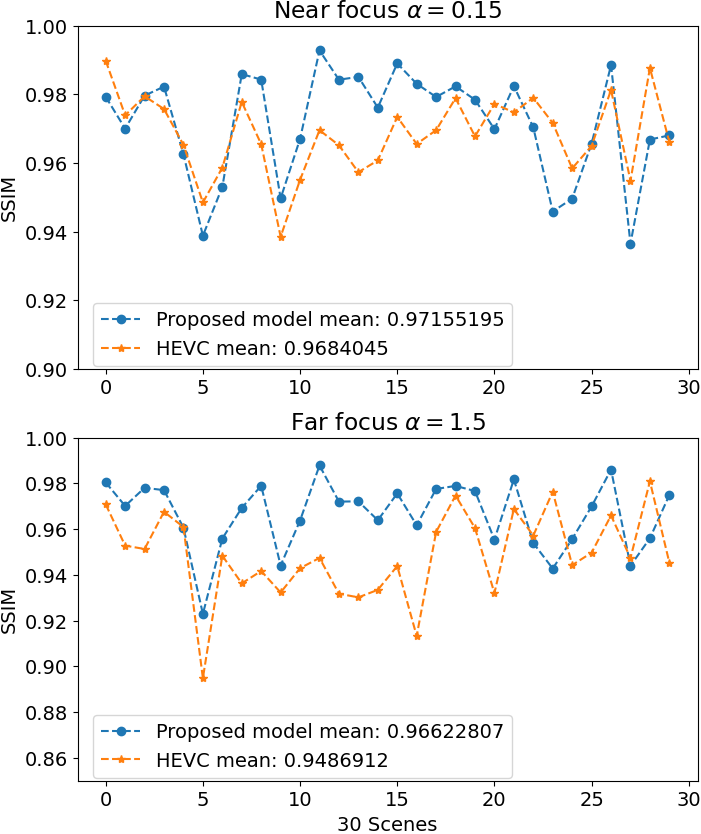}
	\caption{\label{fig:REFSSIM} The SSIM of the near and far focus images extracted from HEVC and our proposed model is calculated and plotted.}
\end{figure}

The PSNR comparison between our model and HEVC over the test set for near and far focus in shown in Fig.~\ref{fig:REFPSNR}. These results show that in some cases our model is superior and for other cases HEVC performs better. The mean PSNR of the HEVC for the test set is greater than ours by 1.3db for the near focus and 0.6db or the far focus.  But for the case of the SSIM metric over the same test set, depicted in Fig.~\ref{fig:REFSSIM}, we can see that in both near and far focuses, our model is performing better. Our model has 0.4\% greater SSIM for near focus and 1.8\% for far focus.

\begin{figure*}
	\centering
	\includegraphics[width=.98\textwidth]{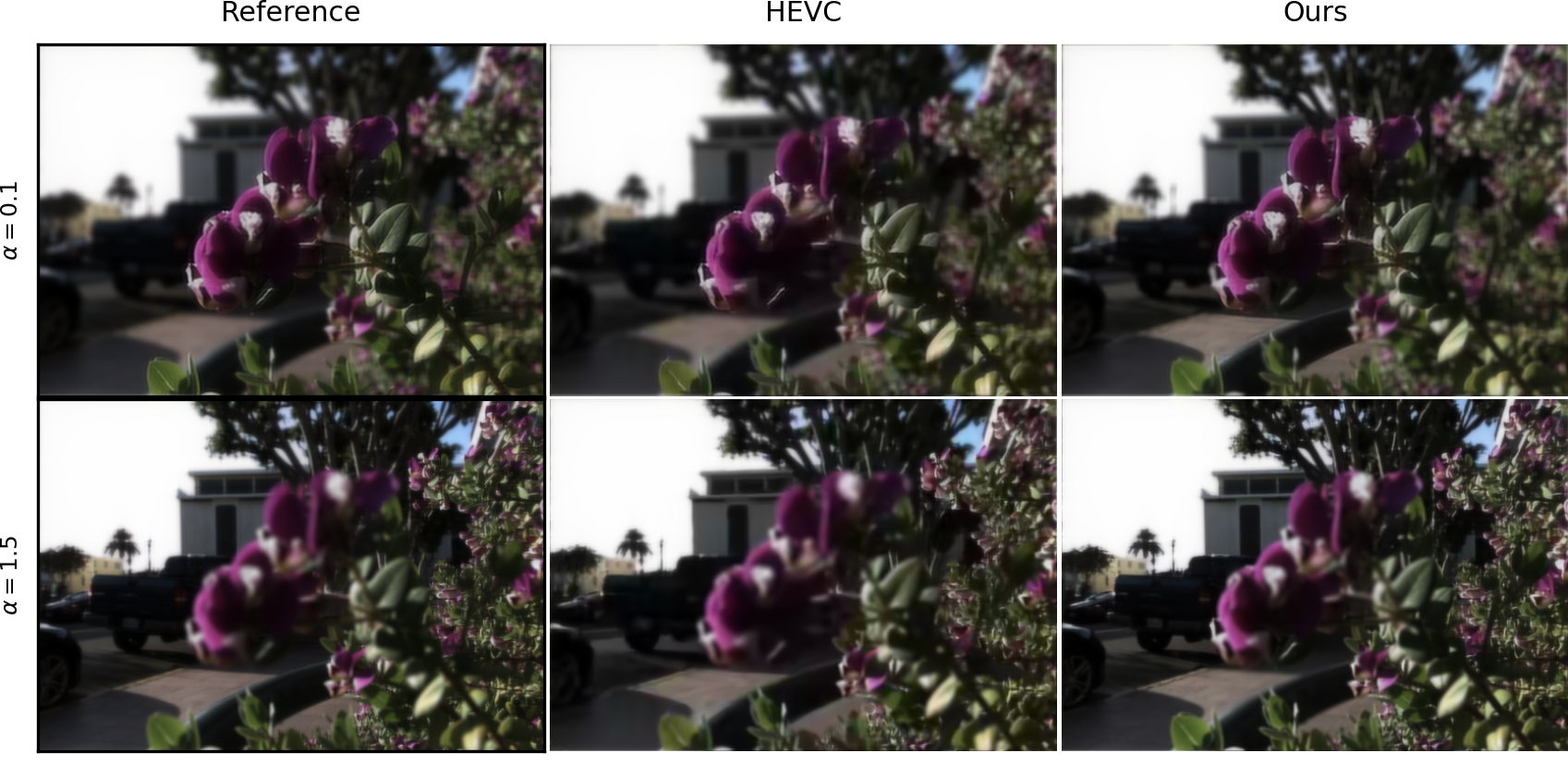}
	\caption{\label{fig:qualitative} Two different focus points of an LF. The refocused DoFs in the top row are focused with $\alpha=0.1$ to the nearest flower, and the bottom row DoFs are  focused at the car with $\alpha=1.5$. The images in the left most column are ground truth images. The middle column shows DoF images extracted from the HEVC reconstructed LF. The rightmost column contains the results from the LF reconstructed by our model.}
\end{figure*}

\begin{figure}
	\centering
	\includegraphics[width=.45\textwidth]{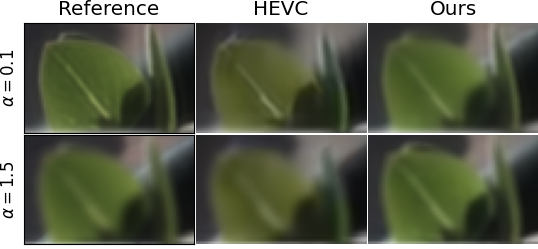}
	\caption{\label{fig:accuracy} In this figure, a small slice from \ref{fig:qualitative} shows HEVC loses more physical information compared to our model. The second leaf just behind the front leaf is not visible in the HEVC reconstructed LF's DoF. }
\end{figure}

\begin{figure}
	\centering
	\includegraphics[width=.45\textwidth]{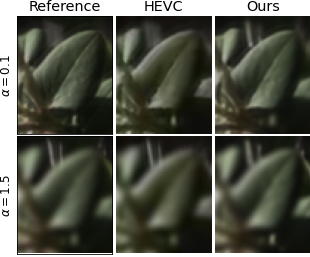}
	\caption{\label{fig:texture} For better texture comparison, a small leaf from the \ref{fig:qualitative} is magnified in this figure.} 
\end{figure}

While the quantitative results look nearly the same between our method and the HEVC compression technique, the reconstructed images show the real differences. Fig.~\ref{fig:qualitative} shows the reconstructed view of a statue from the LF. It is the 25th LF in the 30 Scenes data set. By looking at Fig.~\ref{fig:DOFSSIMPSNR}, we can see that HEVC's MPSNR for this LF is more than 3dB greater than our model. Yet, Fig.~\ref{fig:qualitative} shows that the reconstructed LFs from our model are producing a visually better representation of the ground truth image. The HEVC reconstructed LF generally has more blur all over the image. This blur is, to some extent, caused by severe data loss. Fig.~\ref{fig:accuracy} illustrates another example, where the second leaf behind the front one is not reconstructed in HEVC decompressed LF. Last, we can see more details and better texture in the extracted DoF from our model, demonstrated in Fig.~\ref{fig:texture}. In the refocused images extracted from our model, they have the same depth to the reference image and are refocused to the same focal plane as the ground truth. Images reconstructed from HEVC seem to lose the focal plane, especially in the one focused on the car in Fig.~\ref{fig:qualitative}. Here, it becomes clear that our model is more successful in retaining the LF physical information. On the other hand, this finding indicates that the available quantitative metrics do not tell the whole story in comparing the two LF reconstruction methods. It is worth noting that for quality assessment, the SSIM metric is showing more agreement with the qualitative comparison than PSNR.

\subsection{Speed Gain}
Table \ref{tab:CompTime} shows that the compression time of our proposed method is more than 100 times faster than that of HEVC, on the same computational hardware. This is because our compression pipeline is more efficient, which is just a JPEG algorithm on a fraction of the LF ($1/49$ in our case with 49 views). The HEVC algorithm processes all of the views. 

\begin{table}
	\centering
	\caption{The time takes to compress all 30 LFs in the 30 scenes data set using our method and the HEVC. We can see an speed up of more than 102 times.}
	\label{tab:CompTime}
	\begin{tabular}{l c l}
		\hline
		Method & CUDA &  Comp time (s)  \\\hline
		HEVC & No &  43.53 \\
		JPEG-Hance + Depth-Net & No & \textbf{0.42} \\\hline
	\end{tabular}
	
\end{table}

For decompression, our method is 18 times faster than HEVC; see Table \ref{tab:RecTime}. To give a fair comparison, we used the NVIDIA optimized HEVC codec using the GPU's video decoder. So on the same hardware, this will be the fastest implementation of HEVC.

\begin{table}
	\centering
	\caption{The reconstruction time on all LFs from the 30 scenes data set by our method and HEVC. We can see that our method is 18 times faster in decompressing.}
	\label{tab:RecTime}
	\begin{tabular}{l c l}
		\hline
		Method & CUDA &  Rec time (s)  \\\hline
		HEVC & yes &  0.399 \\
		JPEG-Hance + Depth-Net & yes & \textbf{0.022}\\\hline
	\end{tabular}
\end{table}

Overall, these results indicate that our model is suitable for compressing LF videos with high angular resolution. This is because our method can decompress in near real time inside the GPU without the barrier of transferring high volumes of data from host to GPU. The bandwidth used from host to GPU is equal to the size of only the center view of the LF.
\section{Conclusions}

We have designed a machine-learning assisted LF compression technique. It contains two sequential custom-designed CNNs: JPEG-Hance and Depth-Net. We showed that there is enough information in a highly compressed LF center view to estimate the depth-map of the LF and then use it to reconstruct the whole LF. Also, compression and decompression are faster with our method. We have used the public \emph{Flowers} and \emph{30 Scenes} data sets to conduct our experiment and also to evaluate our model. We have achieved more than 100 times speedup during compression and about 18 times faster reconstruction compared to using HEVC on LF pseudo sequences. Comparing to HEVC, the reconstructed LFs using our method have superior MSSIMs, and they have comparable MPSNRs. Furthermore, the visual quality of the focal plane images reconstructed using our method are superior. For future work, we will try to add other views, with varied relative compression ratios, to further improve the quality of reconstruction. We will also explore options, such as improving the network architecture, other loss functions, larger training data sets, \etc, to enhance the MPSNR. We are also looking forward to deploying our method on an actual LF video to explore the achieved compression ratio and streaming capabilities.

{\small
\bibliographystyle{unsrt}
\bibliography{refs}
}

\end{document}